\documentstyle[twocolumn]{jpsj}
\def\runtitle{Berry phase and quantized Hall effect 
in three-dimension
}
\def\runauthor{Jun {\sc{Goryo}} and Mahito {\sc{\=Kohmoto}}}

\title{%
Berry phase and quantized Hall effect 
in three-dimension
}

\author{%
Jun \textsc{Goryo}\thanks{E-mail; jfg@issp.u-tokyo.ac.jp} and
Mahito \textsc{Kohmoto}\thanks{E-mail; kohmoto@issp.u-tokyo.ac.jp}
}

\inst{%
The Institute for Solid State Physics, University of Tokyo, 
Kashiwa-no-ha, Kashiwa, Chiba, 277-8581, Japan
}

\recdate{\today}

\kword{%
Berry phase, 
quantum Hall effect in three dimension, Bloch electron, topology    
}

\begin{document}
\sloppy

\maketitle


Berry recognized that in quantum mechanics 
a geometrical phase (Berry phase)
arises from the adiabatic process of a system 
around a closed loop in the parameter space\cite{Berry}. 
Thus, it is essentially the Aharonov-Bohm effect\cite{Aharonov-Bohm} 
in the parameter space. It has been shown that the topological formula for 
the Hall conductivity of two-dimensional Bloch electrons\cite{TKNN,Kohmoto-85} 
can be written in terms of the Berry phase\cite{Kohmoto-93}. 
We demonstrate that the above argument is generalized in the 
three dimensional systems. The Berry phase we consider here 
is induced by the adiabatic change of the 
time-dependent vector potential, which represents the electric field as 
well as the magnetic field. Throughout this note, 
we use the unit $\hbar=c=1$. 

We consider Bloch electrons in a uniform magnetic field 
in three dimensions.  
When we consider the ``rational flux'' introduced in 
Ref. \citen{Kohmoto-Halperin-Wu}, we can choose one of the 
primitive vectors of the Bravais lattice ${\bf c}$ 
parallel to the magnetic field, and also we can find the other two 
${\bf a}$ and ${\bf b}$. Then, we take the magnetic field 
\begin{equation}
{\bf B}=\frac{1}{v_0} \frac{2 \pi}{e} \frac{p}{Q} {\bf c}, 
\label{magnetic-field}
\end{equation} 
where $v_0={\bf a}\cdot
({\bf b} \times {\bf c})$, $p$ and $Q$ are integers and $0<Q$. 
The system has the magnetic translation symmetry, and 
in this case, we can choose the magnetic unit cell 
with the primitive vectors ${\bf a}$, $Q {\bf b}$, and ${\bf c}$.     
Let ${\bf G}_a=(2 \pi / v_0) ({\bf b} \times {\bf c})
,~{\bf G}_b=(2 \pi / v_0) ({\bf c} \times {\bf a})$ and 
${\bf G}_c=(2 \pi / v_0) ({\bf a} \times {\bf b})$ stand 
for the primitive vectors of the reciprocal lattice.  
Then, the magnetic Brillouin zone (MBZ) is written as,  
\begin{eqnarray}
{\bf k}=f_1 {\bf G}_a + (f_2 / Q) {\bf G}_b + f_3 {\bf G}_c 
; ~0<f_1, f_2, f_3<1. 
\label{crystal-momentum}
\end{eqnarray}
For the magnetic translation symmetry, 
the wavefunction is written in the Bloch form 
\begin{equation}
\psi^n_{\bf k} ({\bf r}) 
= e^{i {\bf k} \cdot {\bf r}} u^n_{\bf k} ({\bf r}),   
\label{Bloch-form}
\end{equation}
where $u^n_{\bf k}$ has a property 
\begin{equation}
u^n_{\bf k}({\bf r} + {\bf R}^{\prime})=e^{i {\bf R}^{\prime} \cdot {\bf A}}
u^n_{\bf k} ({\bf r}),  
\end{equation}
where ${\bf R}^{\prime}=l {\bf a} + m Q {\bf b} + n {\bf c}$ 
($l,m,n$; integer) and 
${\bf A}$ is the vector potential for the electromagnetic field.

It has been shown that, when the Fermi energy is located in the energy gap,
 the Hall conductivity for the three-dimensional Bloch electrons  
is written as 
\begin{eqnarray}
\sigma_{ij}&=&\frac{e^2}{4 \pi^2}\varepsilon_{ijk} G_k, 
\nonumber\\
{\bf G}&=&-(t_a {\bf G}_a + t_b {\bf G}_b + t_c {\bf G}_c)
\label{sigma_ij}
\end{eqnarray}
where $i,j,k=x,y,z$  
The coefficients $t_a$, $t_b$ and $t_c$ 
are the topological numbers and take integer values\cite{Kohmoto-Halperin-Wu}. 
For example, $t_c$ is 
written as 
\begin{eqnarray}
t_c&=& \sum_{n^{\prime} \leq n} \int^1_0 d f_3 \sigma_c^{(n^{\prime})}(f_3)
\nonumber\\
\sigma_c^{(n)}(f_3)&=&\frac{1}{2 \pi i}\int_{S(f_3)} d^2 k 
\left[\nabla_{\bf k} \times {\bf a}_{n}({\bf k})\right]\cdot \frac{{\bf c}}{|{\bf c}|},  
\label{t_c}
\end{eqnarray}
where we introduce the vector field 
defined by, 
\begin{equation}
{\bf a}_n({\bf k})=< u^n_{\bf k} | \nabla_{\bf k} | u^n_{\bf k} >.  
\end{equation} 
The suffices $n$ and $n^{\prime}$ are 
the band indices and the summation is taken over
 all the bands below the Fermi energy. 
The integral $\int^1_0 d f_3 \int_{S(f_3)} d^2 k$ shows the 
${\bf k}-$integral over the MBZ. 
The expression of $\sigma_c^{(n)}(f_3)$ is the Chern number of
 the fiber bundle defined on $S(f_3)$, i.e. two-torus, whose connection 
is ${\bf a}_n({\bf k})$. It becomes integer and does
 not depend on $f_3$ when the $n$-th band does not cross the Fermi
 level. The detailed discussion can be seen in
 Ref. \citen{Kohmoto-85} and Ref. \citen{Kohmoto-Halperin-Wu}. 
It can be shown that the other integers $t_a$ and $t_b$ are also 
equivalent to the Chern number as the same manner. 

Following Ref. \citen{Kohmoto-93}, we derive the Berry phase of this
 system.  We consider the time-dependent Schr{\"o}dinger equation
\begin{equation}
i \frac{\partial \Psi (t)}{\partial t}=
\left[\frac{1}{2m}(- i {\bf \nabla} + e {\bf A}(t))^2 + U({\bf r}) \right] \Psi (t), 
\label{t-dep-shrodinger}
\end{equation}
where $U({\bf r})=U({\bf r} + l {\bf a} + m {\bf b} + n {\bf c})$ and 
${\bf A}(t)$ is the time-dependent vector potential for the
electromagnetic field. Therefore, we consider not only the uniform magnetic
field Eq. (\ref{magnetic-field}), but also the electric field.  
We use the adiabatic approximation. It means that we introduce a weak
 electric field. According to Berry\cite{Berry} and by using the Bloch
 form Eq. (\ref{Bloch-form}), the wave function of the 
electron for the $n$-th band at time $t$ can be written with a phase 
$\gamma_n (t)$ as 
\begin{equation}
\Psi(t,{\bf r})=
\exp\left[- i \int_0^t dt^{\prime} E_{n\bf k} (t^{\prime})\right]
\exp\left[i \gamma_n (t) \right] 
e^{i {\bf k} \cdot {\bf r}} u^n_{\bf k}(t,{\bf r}),  
\label{wave-function}
\end{equation}
where the function $u^n_{\bf k}(t,{\bf r})$ and $E_{n\bf k} (t)$ are 
obtained by solving the instantaneous eigenvalue equation 
\begin{eqnarray}
H_{\bf k}(t) u^n_{\bf k}(t, {\bf r})
&=&\left[\frac{1}{2 m} (- i {\bf \nabla} + {\bf k} + e {\bf A}(t))^2 
+ U({\bf r})\right] u^n_{\bf k}(t, {\bf r})
\nonumber\\
&=&E_{n\bf k} (t) u^n_{\bf k}(t, {\bf r}). 
\label{eigenvalue}
\end{eqnarray} 
The time-dependent part of the vector potential is written as $- {\bf E} t$. 
Substitute Eq. (\ref{wave-function}) into 
Eq. (\ref{t-dep-shrodinger}) and use Eq. (\ref{eigenvalue}), 
then $\gamma_n(t)$ is obtained as
\begin{eqnarray}  
\gamma_n (t)&=&\int_0^t d t^{\prime} 
\left< u_{\bf k}^{n}(t^{\prime})\left|\frac{\partial}{\partial t^{\prime}}\right|u_{\bf k}^{n} (t^{\prime})\right>  
\nonumber\\
&=&\int_0^t d t^{\prime} 
\left< u_{{\bf k}-e{\bf E}t^{\prime}}^{n}
\left|\frac{\partial}{\partial t^{\prime}}\right|
u_{{\bf k}-e{\bf E}t^{\prime}}^{n}\right>.
\label{gamma_n}
\end{eqnarray}
In order to consider the Berry phase, a Hamiltonian must go around a
 closed loop in a parameter space in adiabatic process. The Hamiltonian
 $H_{\bf k}(t)$ as it is not have this property. However, it is possible 
to compactify it into the magnetic Brillouin zone 
as $H_{f_1 + 1, f_2, f_3}(t) \sim H_{f_1, f_2 + 1, f_3}(t) \sim
 H_{f_1, f_2, f_3 + 1}(t) \sim H_{f_1, f_2, f_3}(t)$ , where 
$f_1$, $f_2$ and $f_3$ parameterize ${\bf k}$ as 
 Eq. (\ref{crystal-momentum}). 
Therefore, in the case that ${\bf E}//{\bf G}_a$,
${\bf E}//{\bf G}_b$ and ${\bf E}//{\bf G}_c$, 
Berry phases are obtained from Eq. (\ref{gamma_n}) as 
\begin{eqnarray}
\Gamma^{(n)}_a (f_2,f_3)&=&
i \int_0^{1} d f_1 {\bf G}_a \cdot {\bf a}_n({\bf k}),  
\nonumber\\
\Gamma^{(n)}_b (f_3,f_1)&=&
i \int_0^{1} d f_2 {\bf G}_b \cdot {\bf a}_n({\bf k}), 
\label{Berry-phase}\\
\Gamma^{(n)}_c (f_1,f_2)&=&
i \int_0^{1} d f_3 {\bf G}_c \cdot {\bf a}_n({\bf k}), 
\nonumber
\end{eqnarray}
respectively. 
By using the Stokes theorem in Eq. (\ref{t_c}), 
$t_c$ can be written in terms of Berry phase as 
\begin{eqnarray}
t_c&=&\frac{1}{2 \pi i}\sum_{n^{\prime} \leq n} \int_0^1 d f_3 \times 
\\
&&
\left[\int_0^1 d f_1 \frac{d}{d f_1} \Gamma^{(n^{\prime})}_b(f_3,f_1) 
- \int_0^1 d f_2 \frac{d}{d f_2} \Gamma^{(n^{\prime})}_a (f_2, f_3) \right],   
\nonumber
\end{eqnarray}
where $\Gamma^{(n^{\prime})}_b (f_3, f_1)$ is the Berry phase for 
the band $n^{\prime}$ below the Fermi energy  
induced by the 
electric field along ${\bf G}_b$. 
The symbol $\sum_{n^{\prime} \leq n}$ denotes that the summation is taken over 
all the bands below the Fermi energy. 
Similarly, we can show that $t_a$ and $t_b$ are also written in terms 
of the Berry phases as 
\begin{eqnarray}
t_a&=&\frac{1}{2 \pi i}\sum_{n^{\prime} \leq n} 
\int_0^1 d f_{1} 
\times 
\label{t-a-3D}\\
&&
\left[\int_0^1 d f_{2} \frac{d}{d f_{2}} 
\Gamma^{(n^{\prime})}_{c}(f_{1},f_{2}) 
- \int_0^1 d f_{3} \frac{d}{d f_{3}} 
\Gamma^{(n^{\prime})}_{b} (f_{3}, f_{1}) \right], 
\nonumber\\
t_b&=&
\frac{1}{2 \pi i}\sum_{n^{\prime} \leq n} 
\int_0^1 d f_{2} \times
\label{t-b-3D}\\
&&
\left[\int_0^1 d f_{3} \frac{d}{d f_{3}} 
\Gamma^{(n^{\prime})}_{a}(f_{2},f_{3}) 
- \int_0^1 d f_{1} \frac{d}{d f_{1}} 
\Gamma^{(n^{\prime})}_{c} (f_{1}, f_{2}) \right]. 
\nonumber
\end{eqnarray}

Therefore, three-dimensional quantized Hall conductivity is 
closely related to the Berry phase. This result is the 3D generalization 
of that which was pointed out in Ref. \citen{Kohmoto-93}. 

The authors thank Masatoshi Sato and Nobuki Maeda for useful
discussions.

\vspace{1cm}


\end{document}